\documentclass[acmsmall,nonacm]{acmart}

\AtBeginDocument{%
  \providecommand\BibTeX{{%
    \normalfont B\kern-0.5em{\scshape i\kern-0.25em b}\kern-0.8em\TeX}}}

\usepackage[utf8]{inputenc}

\usepackage{array}
\newcommand{\PreserveBackslash}[1]{\let\temp=\\#1\let\\=\temp}
\newcolumntype{C}[1]{>{\PreserveBackslash\centering}m{#1}}
\newcolumntype{R}[1]{>{\PreserveBackslash\raggedleft}m{#1}}
\newcolumntype{L}[1]{>{\PreserveBackslash\raggedright}m{#1}}
\usepackage{tabu}
\usepackage{xcolor}
\newcommand{\added}[1]{\textcolor{black}{#1}}

\setcopyright{acmcopyright}
\copyrightyear{2020}
\acmYear{2020}
\acmDOI{10.1145/1122445.1122456}

\begin{document}

%%
%% The "title" command has an optional parameter,
%% allowing the author to define a "short title" to be used in page headers.
\title{Middle-Aged Video Consumers' Beliefs About Algorithmic Recommendations on YouTube}

%%
%% The "author" command and its associated commands are used to define
%% the authors and their affiliations.
%% Of note is the shared affiliation of the first two authors, and the
%% "authornote" and "authornotemark" commands
%% used to denote shared contribution to the research.
\author{Oscar Alvarado}
\authornote{These authors contributed equally to this work.}
\email{oscarluis.alvaradorodriguez@kuleuven.be}
\orcid{0000-0002-3031-9579}
\affiliation{%
 \institution{Department of Computer Science,}
 \institution{KU Leuven}
 \city{Leuven}
 \country{Belgium}
 }

\author{Hendrik Heuer}
\authornotemark[1]
\email{hheuer@uni-bremen.de}
\orcid{0000-0003-1919-9016}
\affiliation{%
 \institution{University of Bremen, Institute for Information Management}
 \city{Bremen}
 \country{Germany}
 }

\author{Vero Vanden Abeele}
\email{vero.vandenabeele@kuleuven.be}
\orcid{0000-0001-5130-8636}
\affiliation{%
 \institution{e-Media Research Lab,}
 \institution{KU Leuven}
 \city{Leuven}
 \country{Belgium}
 }

\author{Andreas Breiter}
\email{abreiter@uni-bremen.de}
\orcid{0000-0002-0577-8685}
\affiliation{%
 \institution{University of Bremen, Institute for Information Management}
 \city{Bremen}
 \country{Germany}
 }

\author{Katrien Verbert}
\email{katrien.verbert@kuleuven.be}
\orcid{0000-0001-6699-7710}
\affiliation{%
 \institution{Department of Computer Science,}
 \institution{KU Leuven}
 \city{Leuven}
 \country{Belgium}
 }

%%
%% By default, the full list of authors will be used in the page
%% headers. Often, this list is too long, and will overlap
%% other information printed in the page headers. This command allows
%% the author to define a more concise list
%% of authors' names for this purpose.
\renewcommand{\shortauthors}{Alvarado and Heuer, et al.}

%%
%% The abstract is a short summary of the work to be presented in the
%% article.
\begin{abstract}
User beliefs about algorithmic systems are constantly co-produced through user interaction and the complex socio-technical systems that generate recommendations. Identifying these beliefs is crucial because they influence how users interact with recommendation algorithms. With no prior work on user beliefs of algorithmic video recommendations, practitioners lack relevant knowledge to improve the user experience of such systems. To address this problem, we conducted semi-structured interviews with middle-aged YouTube video consumers to analyze their user beliefs about the video recommendation system. Our analysis revealed different factors that users believe influence their recommendations. Based on these factors, we identified four groups of user beliefs: Previous Actions, Social Media, Recommender System, and Company Policy. \added{Additionally, we propose a framework to distinguish the four main actors that users believe influence their} video recommendations: the current user, other users, the algorithm, and the organization\added{. This framework provides a new lens to explore design suggestions based on the agency of these four actors. It also exposes a novel aspect previously unexplored: the effect of corporate decisions on the interaction with algorithmic recommendations. While we found that users are aware of the existence of the recommendation system on YouTube, we show that their understanding of this system is limited.}
\end{abstract}

%%
%% The code below is generated by the tool at http://dl.acm.org/ccs.cfm.
%% Please copy and paste the code instead of the example below.
%%
% \begin{CCSXML}
% <ccs2012>
% <concept>
% <concept_id>10003120.10003121</concept_id>
% <concept_desc>Human-centered computing~Human computer interaction (HCI)</concept_desc>
% <concept_significance>500</concept_significance>
% </concept>
% <concept>
% <concept_id>10003120.10003121.10003122.10003334</concept_id>
% <concept_desc>Human-centered computing~User studies</concept_desc>
% <concept_significance>100</concept_significance>
% </concept>
% <concept>
% <concept_id>10002951.10003317.10003347.10003350</concept_id>
% <concept_desc>Information systems~Recommender systems</concept_desc>
% <concept_significance>300</concept_significance>
% </concept>
% </ccs2012>
% \end{CCSXML}

% \ccsdesc[500]{Human-centered computing~Human computer interaction (HCI)}
% \ccsdesc[300]{Information systems~Recommender systems}
% \ccsdesc[100]{Human-centered computing~User studies}

%%
%% Keywords. The author(s) should pick words that accurately describe
%% the work being presented. Separate the keywords with commas.

%%
%% This command processes the author and affiliation and title
%% information and builds the first part of the formatted document.
\maketitle

\section{Introduction}

Recommender systems help users navigate immense \added{number} of movies, songs, news articles, friends, restaurants, and others. As an integral part of platforms like YouTube, Facebook, Netflix, or Amazon~\cite{Jannach:2016:RSB:3013530.2891406}, recommendation systems shape users' everyday experience with information systems~\cite{Willson2017}. Moreover, recommendation systems select and exclude information, defining what is considered legitimate or relevant knowledge~\cite{Gillespie2014} and influencing the behavior and practices of users. Consequently, recommendation systems are part of a highly complex and largely invisible socio-technical system.

As beliefs guide users' behavior, researching them can yield valuable insights into the design of technology. Since user beliefs about algorithmic systems are constantly co-produced and formed during usage, academics need to study them in a specific context with real users. 

In this study, we focus on YouTube, the second most visited website worldwide with billions of users~\cite{YouTube_press_2019}. YouTube's algorithmic recommendation system is responsible for 70\% of the videos \added{consumed on the platform}~\cite{solsman_YouTubes_2018}, and is used by the vast majority of users~(80\%)~\cite{pewresearch_YouTube_2018}. YouTube is \added{also} one of the oldest and most popular social networks with a unique community of video producers and consumers. YouTube's revenue-sharing scheme enables video producers to make a living through their content, which makes \added{them} highly dependent on \added{the} recommendations. \added{This dynamic has} led researchers to focus primarily on the perspective of video producers or YouTubers~\cite{Wu:2019:AGD:3371885.3359321, Bishop2018}. 

\added{Researchers have primarily studied user beliefs about algorithmic recommendations on social media platforms like Facebook~\cite{Rader2015,Eslami2016}, or Twitter~\cite{DeVito2017}. Researchers have not yet attended to user beliefs about video recommendations on YouTube. Such an investigation is important because there are essential differences between the user interfaces of the different platforms. \added{Considering these differences regarding the interface, the interactions, and the system output~\cite{gelman2011concepts, Rader2015}}, a research gap \added{exists regarding user's} beliefs \added{about} video recommendation systems such as YouTube. Moreover, passive video consumers of YouTube have not yet been explored, even though recent media reports have accused YouTube of enabling online ``radicalization''~\cite{nytimes_YouTube_radicalizer_2018,nytimes_chemnitz_2018,roose_YouTubes_2019}. Additionally, this paper focuses on people without formal ICT knowledge who are aged between 37 and 60, complementing similar work that concentrated on people aged 25 or younger~\cite{DeVito2018} or work that addressed broad age ranges from 18 to 64~\cite{Eslami2016}. This middle-age population did not grow up with recommendation systems, which means investigating them provides insights into the perspective of laypeople. Consequently, this demographic is well-suited to understand the varied user beliefs about algorithmic recommendation systems in populations without formal knowledge about algorithmic systems.}

We addressed these research gaps \added{by conducting} 18 semi-structured interviews in three countries with high levels of YouTube usage: Belgium, Costa Rica, and Germany. \added{In line with the research gaps}, we focused on users with three main characteristics: 1)~users who only consume videos, 2)~middle-aged users who did not grow up with social media and algorithmic recommendations, and 3)~users with high education levels but with no formal training \added{in} computer science or related disciplines.

Our analysis identified several influence factors that users recognized in the context of algorithmic recommendations based on the user interface and the output of the system. We grouped these influence factors into four user beliefs about video recommendations on YouTube: Previous Actions, Social Media, Recommender System, and Company Policy. 

We also situate the discovered beliefs \added{in a framework that highlights} the four main actors that \added{video consumers identify as a relevant} influence of the video recommendations: the current user, other users, the algorithm, and the organization. These four actors provide a \added{novel} way to understand the socio-technical context that influences video recommendations. \added{Moreover, this framework extends on prior similar work by adding a new previously unexplored influence factor: the agency of the organization that operates the recommender system. Our results also show a general level of awareness of the recommendation algorithm from the participants, even though the understanding of these recommendation systems remains limited.}

\section{Background}

\added{In this section, we will discuss academic work related to the current study. First, we describe the relevance of studying recommendation systems because of their social implications and consequences towards their users. Second, we highlight previous work on how users perceive and experience these systems and related design proposals. Third, we explore previous investigations of algorithmic beliefs and related research. Finally, we present official public sources that describe how YouTube's recommendation system works.} 

\added{In this paper, we use the term recommendation system (RS) to refer to systems based on machine learning, collaborative filtering, or other user-content based recommendation strategies. Since we center our investigation on users and their understanding and experiences about these recommendations systems, we highlight specific technical implementation details only when they relate to this topic. In addition to that, the participants in our investigation did not differentiate between concepts like collaborative filtering, machine learning or neural networks, neither explicitly nor implicitly.}

\subsection{The Relevance of \added{Algorithmic Recommendation Systems}}

Different academic efforts have analyzed the relevance and implications of \added{RS} for users and societies. \added{Gillespie~\cite{Gillespie2014}, for instance, discusses \textit{public relevance algorithms} that select or exclude information, infer or anticipate user information, define what is relevant or legitimate knowledge, flaunt impartiality without human mediation, provoke changes in the behavior and practices of users, and produce calculated publics~\cite{Gillespie2014}}. Similarly, Cosley et al. argue that specific \added{RS can} affect the opinions of people about the content they recommend~\cite{cosley_is_2003}. Likewise, Willson and Beer suggest studying algorithms that work semi-autonomously and exert power with no supervision from human counterparts ~\cite{Willson2017, Beer2017}.

\added{This body of work highlights the relevance of recommendation systems and their implications for users and society. YouTube's RS is also \textit{publicly relevant} due to its potential to guide users' opinions and take decisions over the information that the system selects.}  

\added{A} large body of academic work centers on the relevance of algorithms that recommend cultural content. Striphas, for instance, presents the concept of algorithmic culture as ``the enfolding of human thought, conduct, organization and expression into the logic of big data and large-scale computation''~\cite{Striphas2015}. Morris explains that \added{recommendation systems} define the current relationship between cultural products and consumers, impacting culture management, and consumption~\cite{Morris2015}. Additionally, Prey analyzes how \added{recommender systems} define their audiences, arguing that personalized media \textit{pretends} to define distinct preferences of users: ``there are no individuals, but only ways of seeing people as individuals''~\cite{Prey2017}. Likewise, Gillespie explains how ``trending'' algorithms define specific audiences based on profiles~\cite{Gillespie2016}, and how these profiles are becoming a source of cultural concern themselves. Furthermore, Rieder et al. determined the extent to which the results provided by the recommender algorithm on YouTube are based on popularity but also ``specific vernaculars'' such as the video issue date and YouTube's definition of novel videos~\cite{Rieder2018}.

\added{In contrast to other social media platforms such as Facebook or Twitter that are focused on sharing a variety of content, this work shows how a platform such as YouTube's RS centers on recommending video content as particular cultural products, organizing users in specific niches, and determining trending content. Besides differentiating between YouTube and other social media, this dynamic also affects the distribution, the consumption, and the relevancy of cultural products on this platform.} 

\added{Recommendation systems} benefit both the platform providers and users by increasing usage time and improving the user experience. Despite these benefits, scholarly efforts documented various problems associated with the application of these algorithms. For example, Bozdag describes the many layers of bias that could affect algorithmic filtering and personalization~\cite{Bozdag2013}. Moreover, Mittelstadt et al. provide an extensive survey of the ethical issues associated with algorithms such as unjustified actions, opacity, bias, discrimination, challenges for user autonomy, privacy, and moral responsibility~\cite{Mittelstadt2016}.

\added{Additionally, media reports} suggest that video recommender systems promote extreme videos that affect users' opinions on a topic~\cite{roose_YouTubes_2019}, which can expose users to extreme ideas about politics and other social issues. For instance, Lewis showed how political organizations build audiences and sell their content, thus enabling far-right influencers~\cite{Lewis2018}.

\added{These works highlight the relevance of YouTube's RS, a platform that poses the risk of creating filter bubbles and the risk of exposing users to biased information. These characteristics indicate that YouTube's RS is different from other social media platforms such as Facebook or Twitter. Therefore, it is essential to explore users' understanding and perceptions of YouTube's RS to find clues as to how these risks exist and how to reduce them.} 

\subsection{User Experience of Algorithms}

The experience that users have with algorithms attracted much interest in recent years. Oh et al. propose \textit{algorithmic experience} as a: ``new stream of research on user experience''~\cite{Oh2017} that considers the constant relationship between users and algorithms. For Alvarado and Waern, \textit{algorithmic experience} (AX) is an: ``analytic framing for making the interaction with and experience of algorithms explicit''~\cite{Alvarado2018}. In a subsequent study, Alvarado et al. proposed a framework for AX dedicated to movie recommender algorithms that distinguish different design areas like profiling transparency and management, algorithmic awareness, user-control, and algorithmic social practices remembering~\cite{Alvarado2019}.

Other studies \added{examined} the level of awareness of algorithmic systems among users. Hamilton et al. investigated the role of algorithms and filters in algorithmic news curation~\cite{hamilton_path_2014}. In their sample, less than 25\% of regular Facebook users were aware that their feeds are curated or filtered. Similarly, Eslami et al. report that in their experiment, less than half (37.5\%) of the participants are aware of the News Feed curation algorithm's existence \cite{Eslami2015}. Eslami et al. also found that users are upset when the curation algorithm does not show posts by close friends and family. Surprisingly, users even believed that their friends intentionally chose not to show them these posts. The study also showed that users becoming aware of algorithmic curation could provoke angry feelings about not seeing posts from close friends or family members.

Wu et al. investigated how content creators on YouTube craft algorithmic personas based on their experience with the \added{RS} on YouTube~\cite{Wu:2019:AGD:3371885.3359321}. They identified three algorithmic personas on YouTube that creators distinguish: the Agent, the Gatekeeper, and the Drug Dealer. While users saw the Agent as a friend that procures employment, the Gatekeeper was a persona that users tried to bri\added{b}e to get their content viewed. The Drug Dealer, on the other hand, was focused on keeping viewers addicted to the platform. 

\added{Similarly, Pires et al. explored the practices and metaphors of teens that use YouTube~\cite{doi:10.1080/1369118X.2019.1672766}. They found that teens use different metaphors to describe YouTube. These metaphors include YouTube as a search engine, a Smart TV, a distribution channel, a co-creation space, and an informal learning space. However, they did not explore the role that the algorithmic recommendation system has in this context and for this group of users.}

\added{All of this work invites to explore the awareness and understanding of YouTube's RS within a specific subset of video consumers. In contrast to these studies, we found a comparatively high level of awareness of the RS on YouTube on video consumers, with limitations regarding the understanding of the inner working of the system}.

\subsection{Mental Models, Folk Theories, and Algorithmic Beliefs}

Research on the different understandings that users possess about algorithms and their processes has a long tradition. Academics have explored those understandings through different theoretical framings such as mental models, folk theories, or user beliefs. 

According to Norman, people formulate mental models of a system through interaction~\cite{Norman:1987:OMM:58076.58097}. Such mental models are, therefore, evolving ``naturally'' over time. They are incomplete, and their ability to ``run them'', i.e., to predict system behavior, is limited. Norman also stated that these models are not technically accurate. However, they have to be functional, which means that users continuously modify mental models to get a workable result. In addition to that, a user's technical background or previous experience with similar systems can constrain the mental model. 

In our current study, we do \added{not} depart from the concept of mental models for several reasons. First, we are studying users who are not directly interacting with YouTube's \added{RS}, which means that we cannot investigate users' ability to ``run'' their mental models. Second, since the public understanding of YouTube's system is limited, we can not investigate the accuracy of a user's mental model because there is no conceptual model of YouTube that we can compare the user's mental model \added{to}. Even those who train these systems cannot offer detailed or complete explanations, especially not \added{for} neural networks~\cite{hamilton_path_2014, Mitchell:1997:ML:541177, Goodfellow-et-al-2016}.

Motivated by the problems of investigating mental models, researchers developed folk theories and user beliefs as theoretical framings to study complex socio-technical systems like \added{recommender systems}. French and Hancock define the term folk theory as a ``person's intuitive, causal explanation about a system that guides their thoughts, beliefs, and actions with that system''~\cite{French2017}. Similarly, DeVito et al. explored different social media platforms to describe how folk theories are built, identifying different sources of information that build them~\cite{DeVito2018}. DeVito et al. also described algorithmic resistance in social media, portraying all understandings related to the insertion of a filtering algorithm on Twitter~\cite{DeVito2017}. Eslami et al. identified several folk theories of Facebook's News Feed~\cite{Eslami2016}. \added{Siles et al. found two main folk theories in Spotify: one that personifies the system as a social being that is providing recommendations based on surveillance, and one that considers Spotify as a computational machine trained by users~\cite{doi:10.1177/2053951720923377}.} 

This investigation \added{examines user beliefs about} the video recommendation system on YouTube. We adopt the term user beliefs from Rader and Gray, who investigated \added{user} beliefs \added{about} algorithmic curation on Facebook~\cite{Rader2015}. In their study, they identified six beliefs. First, passive consumption, related to the belief that there is no algorithmic curation. Second, producer privacy, as the belief that the algorithmic curation exists because friends define the audience for their posts, excluding specific people to access them. Third, consumer preferences, the belief that the News Feed does not show what the users prefer to see without direct intervention. Fourth, missed posts, the belief that blames the News Feed as the agent responsible for missing specific posts. Fifth, violating expectations, the belief caused by patterns or regularities. Finally, speculating about the algorithms, which connects to the belief that there is an algorithm that filters the posts.

\added{Considering the relevance of the RS on YouTube and its differences to previously studied social media platforms, video consumers' beliefs about the RS in YouTube remain a research gap. This paper addresses this research gap, focusing on middle-aged users, a noteworthy population who did not grow up with these technologies.}

\subsection{The Recommendation Algorithm on YouTube}

Media and academic reports increasingly portray YouTube's \added{RS} as a catalyst for filter bubbles and online radicalization~\cite{roose_YouTubes_2019, Lewis2018,nytimes_YouTube_radicalizer_2018,buzzfeed_las_vegas_2017,nytimes_chemnitz_2018}. However, there is little systematic research \added{to} support these claims. On the contrary, the consensus that recommendation systems are central to the promotion of political content is challenged by Munger and Phillips~\cite{munger2019supply}. While Munger and Phillips acknowledge the algorithm on YouTube as one part of a complex socio-technical system that pushes extreme and radical content, \added{their perspective on online ``radicalization'' focuses on} content created by fringe political actors. Such actors target disaffected individuals who search for sociality online and are alienated from mainstream media by their content. Munger and Phillips argue that YouTube has particular affordances that foster content creation for such fringe political actors, which implies that the \added{RS} on YouTube is only one part of a large and complex socio-technical system. Therefore, studying how users understand these algorithmic systems is crucial \added{to describe} the complexity of the broader socio-technical system. 

In this paper, we situate the user beliefs of our participants by comparing them to statements that describe how YouTube's recommender algorithm works. \added{The system uses personalization and performance to define the selection of the recommended videos.} YouTube states that the recommender algorithm includes ``videos that are news, watched by similar viewers, or from your subscriptions''~\cite{YouTube_creators_home}. YouTube also cites video titles, thumbnails, descriptions, and ``how other viewers seem to be enjoying [the video]''~\cite{YouTube_creators_algorithm} as factors that influence their recommendations. \added{Additionally, }YouTube \added{considers} how much time a person spends watching a video, whether users clicked on the like/dislike buttons and the number of comments a video has. To gain deeper insights into what data is YouTube potentially using, we reviewed YouTube's publicly available application programming interface \added{(API)}. YouTube's API cites the following core view and engagement metrics: the number of views, the percentage of viewers that the system logged in when watching the video or playlist, the number of minutes that users watched, the average length of video playbacks and the number of comments, likes, dislikes, and shares~\cite{googledev_metrics_2019}. 

\added{We also reviewed the limited available academic work related to this topic. An early version of the RS} on YouTube, as described by Davidson et al. in 2010, is based on association rule mining, which applies collaborative filtering to find unseen videos based on the activity of other users~\cite{davidson2010YouTube,hipp2000algorithms}. The similarity between videos is defined based on how often videos are co-watched. Co-watched, in this context, refers to whether the same user consumed two videos within \added{24 hours}. Davidson et al. list two classes of data sources that \added{were} used by the recommendations on YouTube: 1)~content data, including the raw video streams and video metadata like titles and descriptions, and 2)~user activity data, which can either be explicit like video ratings, liking and subscribing, or implicit like starting to watch a video or watching a large portion of a video~\cite{davidson2010YouTube}. A \added{more recent} publication by Covington et al. \added{in 2016} suggests that the \added{RS} on YouTube \added{was based on a machine learning system that} uses two neural networks. \added{Here, o}ne neural network generates candidates based on the videos watched, search query tokens, and demographics, and one neural network ranks the videos~\cite{covington2016deep}. \added{However, it remains unclear whether YouTube still uses these systems in practice.}

These statements present official and academic explanations about the inner workings of the \added{RS} on YouTube. Unfortunately, the company does not offer more detail on the \added{calculation strategies} they use. \added{YouTube also does not describe which machine learning technique is applied, if they base the system on collaborative filtering or neural networks, or a combination of techniques.} Nevertheless, this information serves as a reference to compare the level of awareness that users have regarding the \added{RS} on YouTube \added{with the different technical approaches that are likely to be applied}. 

\section{Method}

Previous work that studied user beliefs and experiences of algorithmic systems applied a variety of different methods, ranging from interviews~\cite{DeVito2018, Eslami2016}\added{,} survey work~\cite{Bernstein:2013:QIA:2470654.2470658}, to the analysis of public tweets~\cite{DeVito2017}. \added{Our investigation is focused on YouTube, a platform that has essential interaction differences in comparison to other platforms like Facebook and Twitter, revealing an exciting research gap.}

\added{For instance, recommendations on Facebook and Twitter are only available for active user accounts. Video recommendations on YouTube can be experienced without logging in the system, or without explicitly connecting with channels. An additional difference in YouTube is that a user usually \textit{subscribes} to channels, creating a uni-directional relationship between a video creator and a video consumer. In contrast, other social media create bi-directional connections (or ``friendships''). Here, both parties can be content creators and consumers. Likewise, the user feeds on Facebook and Twitter are not labeled as recommendations, while YouTube clearly labels them as such.}

For this study, we conducted semi-structured interviews with 18 middle-aged \added{YouTube video consumers}. Semi-structured interviews enabled us to ensure that we covered the most important questions while allowing participants to express their views in their terms, ensuring both depth and breadth. 
 
\subsection{Recruitment and Participants}

We used a non-probabilistic sampling aimed at maximizing diversity. The two first authors recruited participants in more than one country to gain a diverse perspective \added{on} how users reason about the recommendations on YouTube. We wanted participants to be familiar with YouTube. Therefore, we selected countries with high levels of YouTube usage: Costa Rica, Belgium, and Germany. Costa Rica has the highest (59\%) YouTube usage among Latin American countries and a high overall social media usage~\cite{latinobarometro_report_2018}. Germany and Belgium have a high level of YouTube usage (69\%) \added{among} European countries~\cite{we_are_social_digital_2018}. Recruiting participants from the Global South allowed us to not only represent users from countries in the Global North, which are frequently subjects of such studies. That said, even though we recruited participants from different countries, our study is not focused on comparing cultural differences towards algorithm beliefs. The main goal was to gather a broad range of individual perspectives on what factors influence recommendations, including diverse voices that \added{are frequently not represented} in such investigations.

\added{Since we selected participants from different countries, we ensured homogeneity among the participants by controlling for other possible socio-demographic characteristics. We then recruited YouTube video consumers who had at least a university degree to gather data within similar socio-economical contexts.}

Prior research showed that a user's technical background constraints his or her mental models~\cite{Norman:1987:OMM:58076.58097}. Previous research has \added{also} documented how different levels of technical knowledge influence the formation of user beliefs and folk theories~\cite{DeVito2018}. Users with better web skills, for instance, formed their folk theories differently than those with less technical abilities. We controlled \added{for} these factors \added{in all the three countries} by recruiting YouTube users without a background in technology or high ICT literacy. Participants were required not to have formal training or work experience in computer science, programming, or related fields. This decision allowed us to make sure that users' prior experience with such systems and technical backgrounds were \added{comparable}.

We recruited participants \added{aged} 35 or older. \added{Besides seeking homogeneity among the participants, this sampling criteria were defined for three more reasons. First, since users with low ICT literacy also delimited the recruitment, this middle-age sampling improved our chances of addressing a population who} did not grow up with social media or algorithmic systems. \added{Second, researchers had not exclusively addressed middle-aged users in similar studies (e.g., ~\cite{DeVito2018,Eslami2016, Wu:2019:AGD:3371885.3359321}), allowing our study to address this gap in the research \added{and to provide} evidence that previous studies and our results \added{can} be generalized without age concerns. Third, }this delimitation allowed us to include a population that is usually not represented in this kind of study. 

Participants were required to have used YouTube for more than a year and at least once per week. \added{This} ensure\added{d} that they had sufficient experience with the platform. 

Finally, as a way to center the study on video consumers, we intentionally excluded users who considered themselves to be YouTube producers. \added{We also excluded users who have} a YouTube channel or who uploaded videos in the last two months before the investigation. 

We performed \added{the} recruitment through flyers and online bulletin \added{boards}. The final sampling resulted in a gender-balanced (18 total, eight female) group of participants from three countries: six Belgians~(P2, P5, P8, P11, P13, and P16), six Germans~(P1, P4, P7, P10, P15, and P18), and six Costa Ricans~(P3, P6, P9, P12, P14, and P17). \added{Native speakers} conducted both the interviews in Germany and Costa Rica. \added{The interviews} in Belgium \added{were conducted in English. B}oth participants and interviewer were non-native speakers. The mean age of participants was 43.88~(SD=7.04). Twelve participants were between 37 and 43 years old, and three participants were between 47 or 50. The remaining three participants were older than 50. The oldest participant was 60 years old. \added{The} sampling resulted in a highly educated sample: 50\% of the participants had a Bachelor's degree as the highest degree obtained, three participants a Master's degree, and two a Ph.D. The two first authors conducted all interviews between January and May 2019.

\subsection{Procedure During the Interviews}

The two first authors conducted every interview in three phases. We combined a sensitizing exercise in the first phase, a non-biased method in the second phase, and a suggestive method in the third phase. The three methods complemented each other and allowed us to get a holistic and diverse perspective on the user beliefs around the recommendation algorithm on YouTube. \added{All} interviews were audio-recorded and transcribed.

In this first phase, participants answered a structured questionnaire that covered demographic data, their weekly YouTube usage, whether they knew about the existence of the recommendation system, and how much control they think they had over the system. \added{We asked participants whether they knew that YouTube has video recommendations. To verify whether they really knew about the recommendations, participants had to point out the recommendations in the interface. We also asked participants how frequently they consumed the recommendations on the landing page or the recommendations that appear next to each video. We} asked these questions to sensitize the participant and foster a reflection \added{on} the recommendation system. Additionally, each participant \added{was invited} to access YouTube with a computer or a tablet and to review the interface. During the entire interview, participants were able to check their recommendations and the platform to confirm their beliefs.

In the second phase, we invited the participants to draw a concept map while explaining all aspects that they considered as influences for their recommendations on YouTube. Concept maps are a structured way of organizing and representing knowledge that visualizes concepts and the relationships between concepts~\cite{novak2010learning,novak2006theory}. The primary motivation of the concept maps was to elicit reactions from the interviewees and to provoke structural and critical thinking. 

The third phase started after the participants stated that they could not come up with more possible influences. In this phase, interviewers presented possible influence factors. \added{The two first authors derived these} influence factors from official statements about the recommendation system on YouTube \added{described in the related work. These factors} included 1)~channel subscriptions, 2)~user location, 3)~likes, 4)~sharing of videos, and 5)~ comments~\cite{YouTube_creators_algorithm,davidson2010YouTube,covington2016deep}. 

\added{While the} first phase made sure that all participants knew \added{about} the scope of our questions, the second phase mitigated priming and framing effects by allowing users to freely discuss the influence factors they believed in without interference from the interviewers. These open questions during the second phase also enabled us to gather a broader perspective on the different user beliefs, capturing those that come naturally to users. Finally, the third phase allowed us to further contextualize the results, by allowing participants to agree or disagree with the suggestions provided by the interviewers, an aspect that is usually not measured in previous studies~\cite{DeVito2017,Eslami2016,Rader2015}. Moreover, the third phase allowed participants to express other beliefs that they could have forgotten during the second phase.

\subsection{Analysis}

The interviews were analyzed using thematic analysis, a ``foundational method for qualitative analysis'' used for identifying and reporting themes within a data set. It ``provides a flexible and useful research tool, applicable for many theoretical and epistemological approaches''~\cite{Braun2006}. We performed an iterative and collaborative process of inductive coding with weekly meetings in which we discussed the themes and concepts relevant to our investigation. 

Following the methodology, the two first authors of this study steadily moved back and forward between the entire data set, reviewing the transcripts of the interviews several times. After this, both first authors independently wrote down initial codes. The two first authors grouped those codes into potential themes. Both first authors repeatedly reviewed, debated, and solved disagreements during meetings in several iterations. After this phase, all authors reviewed a preliminary set of themes, \added{leading to} a definitive set of themes \added{reported}. Finally, the two first authors gave names to the themes reported in the following section. To present the results while maintaining anonymity, we refer to participants as P(N), where N is a participant from 1 to 18.

\section{Middle-Aged Video Consumers' Beliefs About Video Recommendations on YouTube}

This section provides an overview of the user beliefs and influence factors that middle-aged video consumers have about video recommendations on YouTube. The different user beliefs that we discovered \added{are} based on the thematic analysis. The thematic analysis also yielded a set of influence factors that users relate to the identified user beliefs.

\added{In general, there is no explicit agreement on what factors different middle-aged video consumers believe influences their recommendations. The analysis found no dominant factors and no differences due to gender, education, or country.}

\added{While users showed a high awareness of YouTube's recommender system, their conceptions of the recommendation system remain elusive and poorly specified. This is reflected in the way they talked about the recommendations on YouTube. Users commonly referred to the recommendation system as ``the system'' (six mentions) or ``the algorithm'' (eight mentions). The term ``recommender system'' is only mentioned once. Technical terms like ``collaborative filtering'', ``machine learning'', or ``neural network'' were never used by the participants.}

\added{Table~\ref{tab:beliefs} provides an overview of the different influence factors and user beliefs that this investigation uncovered. In the following, we will describe the different influence factors grouped by the general themes they belong to.}

\renewcommand{\arraystretch}{1.5} % 1.5 also ok
\begin{table}[]
%\centering
\caption{Middle-aged Users’ Beliefs about YouTube’s Recommendation Algorithm grouped the seven most salient groups distinguished by the respondents.}
%\begin{tabular}{lp{6cm}l}
\begin{tabu} to \textwidth {X[2cm,p,l]X[6cm,p,l]X[3.5cm,p,l]}
\toprule
\textit{Beliefs}                     & \textit{Description}         & \textit{Influence Factors}                 \\
\midrule
\textbf{Previous Actions} & This user belief refers to the previous actions of the current user. Respondents believe that the videos a user watched, the videos s/he endorsed or opposed, the keywords the user searched for, the accounts s/he asymmetrically followed, and the videos s/he shared influence the recommendations. & $\bullet$~My Watch History\newline $\bullet$~My Search History\newline $\bullet$~My User Subscriptions\newline $\bullet$~My Likes \& Dislikes\newline $\bullet$~My Comments\newline $\bullet$~My Shared Videos \\
\textbf{Social Media} & This belief references the influence of the activity of other users. This includes statistics about the popularity of a video, other users showing their virtual endorsement or opposition, as well as other users discussing videos in the comments section. & $\bullet$~Others' Viewing Activity\newline $\bullet$~Others' Likes \& Dislikes\newline $\bullet$~Others' Comments \\
\textbf{Recommender System} & With this belief, the respondents refer to the actions of the algorithm that recommends the videos. Influence factors include how the similarity between users and the similarity between videos is computed. Respondents also believe that when they watch something and where they watch it is taken into account. & $\bullet$~Who is similar?\newline $\bullet$~What is similar?\newline $\bullet$~When do I watch?\newline $\bullet$~Where do I watch? \\
\textbf{Company Policy} & This belief relates to the actions of the organization that runs the system. This includes the idea that some recommendations are paid for, the possible influence of data-sharing practices between different companies, as well as psychological experts that are hired to keep users on the platform and to increase profit. & $\bullet$~Paid Recommendations\newline $\bullet$~Third-Party Data-Sharing\newline $\bullet$~Psychological Experts \\
\bottomrule
%\end{tabular}%
\end{tabu}
\label{tab:beliefs}
\end{table}

\subsection{Previous Actions Beliefs}

This group of user beliefs relates to the actions a user performs \added{on} the platform. In this case, various influence factors are taken into account to influence the video recommendations. 

\subsubsection{``My Watch History'' Influence}

Participants believed that their previous watch history influences their recommendations. P9 said: \textit{``I have noticed that they offer videos related to what I have watched [previously]''}. P14 explained:

\begin{quote}
When I enter to see a video, and I go to the recommendations made by YouTube, I think it starts making the statistic~[sic], counting in what categories I visit to make more emphasis on those recommendations for that topic. 
\end{quote}

Similarly, P11 and P16 expressed: \textit{``I think the biggest chunk [of recommendations] is [from] my previous watch list''}. P4 believed that YouTube tracks this watch history using \textit{``some cookies that identify me as a user''}.

\added{Interestingly, participants did not mention the time a user spends watching a specific video. This influence, as explained in the background section~\cite{googledev_metrics_2019}, is a reason that could make a difference in the recommendations.}

\subsubsection{``My Search History'' Influence}

Participants explained that recommendations are offered depending on what the user types into the video search bar. P1, P14, P17, and P9 used similar examples to explain their rationale: \textit{``I can look for how to tune a guitar, and then it starts appearing content related to singers that play guitar''}. Similarly, P2 said: \textit{``I think actually it's like a topic in the search, in your search bar''}. Moreover, P13 explained that his recommendations depend on the different languages he uses to type in the search bar.

Additionally, participants think that YouTube keeps track of the search terms they have used in the past. For example, P4 stated: \textit{``There are individual terms on the page that are displayed to me based on my last searches''}. Also, P14 said: \textit{``YouTube starts noticing that my visits are recurrent in this topic, the next time I enter [the platform], the categories start to be distributed as suggestions''}.

\added{This influence, in particular, shows that participants have an impression of a ``tracking practice''. Even if searching for content does not produce direct ``hits'' for selected videos, participants consider typing in the search bar as something that the system will use to infer their interests and produce future recommendations.}

\subsubsection{``My User Subscriptions'' Influence}

Participants also mentioned how subscriptions on YouTube influence the recommendations~(P3, P5, P7, P16, P17). Surprisingly, the Subscription influence as an explicit way of expressing interest in specific videos was not brought up by the participants without assistance. However, participants agreed on the influence of subscriptions when they were asked about them by the researchers during Phase 3.

P3 believes that being subscribed to a channel would be \textit{``even more reason to take this video into account''}. P5 observed that after subscribing to a channel about guitar instructions, he got recommendations about similar teachers. Likewise, P7 reported that subscribing to a channel of a particular subject led to related suggestions. Similarly, P17 expressed that being subscribed to a channel \textit{``means that YouTube will suggest from that channel because it is from the user's interest''}. P16 even estimated that the most recent uploads of his subscriptions \added{lead to} 60\% of his recommendations.

\added{Interestingly,} participants had different beliefs about how subscriptions work. Statements by P5 and P7 suggest that subscribing to a channel promotes recommendations on a specific topic. P17 thinks that subscribing influences the system to focus only on recommendations from that specific channel. P10 stated that the number of subscriptions a channel has would increase the chances of appearing in the recommendations. 

\added{These expressions of the participants provide a clear example of how superficial their level of awareness about the recommendations is. They also highlight the role that subscriptions play for the recommender system. While users recall that subscribing influence the recommendations, they cannot precisely explain how it happens, creating various beliefs that fill this gap.}

\subsubsection{``My Likes \& Dislikes'' Influence}

Likes and Dislikes are another influence that users regarded as a plausible influence for video recommendations. This influence was predominantly discussed based on the individual actions of users that could affect their recommendations.

For instance, P2 argued that: \textit{``It seems logical that if you like a video, he [YouTube] will say: `You want to see more?', `Of course.' `Here they are.'''}. Users agreed that likes are \textit{``very much involved''}~(P4) in the recommendation process and that more likes lead to a higher ranking~(P1). Similarly, P14, P5, and P2 described likes as an explicit acknowledgment of interest by a user. 

\subsubsection{``My Comments'' Influences}

Participants regarded the comments a user writes on videos as a factor that influences recommendations. Even though a large number of participants agreed that commenting influences video recommendations~(P6, P8, P7, P2, P10), they did not mention it without the researchers' assistance during Phase 3. This behavior implies that this influence is plausible but not intuitive for the users. 

P2 described the influence of commenting as automatically making \textit{``a link''} to the user's interests. P6 suggests that commenting \textit{``can intervene because that reflects a particular interest''} of the user. P10 regarded comments as an influence but stated that this aspect is less important than sharing a video. 

\added{Overall, while users did recognize the influence that commenting could have on the recommendations, they did not elaborate on how the algorithmic recommendations take them into account. Additionally, users showed no awareness of how YouTube uses natural language processing or sentiment analysis techniques to extract relevant information from comments and how this could influence recommendations.}

\subsubsection{``My Shared Videos'' Influence}

Users also believed that sharing videos influences the video recommendations on YouTube~(P4, P6, and P8). P10 described sharing as \textit{``hav[ing] a mission and want[ing] to convince others that this is also interesting''}. 

While participants agreed that sharing has an influence, they rarely elaborated on how it influences the recommendations. P4 said that sharing \textit{``has consequences''}, but she suggested that it might be less critical than, for instance, comments. P8 argued that sharing a video means \textit{``much bigger revenue from that video because other people start watching it''}.

\added{Interestingly, users consider sharing a video as a factor that influences the recommendations, even if YouTube does not display the number of shares of a video in its interface. Therefore, it seems this influence is mostly related to an action by the current user that indicates an interest in a specific video that will later affect the recommendations.}

\subsection{Social Media Beliefs}

Complex socio-technical systems like YouTube provide a variety of social navigation features that guide users through the information space by visualizing the activity of others and allowing users to make decisions based on the decisions of others~\cite{dourish1994running}. In this section, we group all responses that are related to the social media \added{features} of YouTube and its influence \added{on} the video recommendations. \added{This group of influences indicates that participants perceive YouTube as a social media platform, rather than a passive video consumption service.} 

\subsubsection{The ``Others' Viewing Activity'' Influence}

Respondents believe that the number of views of a video influences the recommendations produced by the algorithm. P1, for instance, expressed that those videos that \textit{``are more successful appear first''}. P10 thought \textit{``that the click rate might have an influence on that. So how often other people have already clicked on this video''}. 

Although, \added{P6 believed that} this dynamic could \added{lead to} negative experiences with the platform. P6 explained this as follows:

\begin{quote}
There are other people using the platform, so these people are creating a tendency, maybe a song that I do not like but they do listen to it frequently, so probably YouTube will believe that it will possibly also be likable for me and it [YouTube's recommendation system] will include that song inside my recommendations. [...] That is why the recommendations will have videos that I really do not like. There are videos that I really hate, like animal mistreat or related, but there are many people that watch those kinds of contents [...]. If all those people watch those videos, probably YouTube is going to recommend those videos.
\end{quote}

\added{This previous quote shows that users would prefer to control the influence that other people have on their recommendations. We found both positive and negative perspectives on this influence.}

\subsubsection{The ``Others' Likes \& Dislikes'' Influence}

Participants considered the influence that the likes and dislikes of other users have on the recommendations~(P1, P5, and P18). However, participants did not bring up the influence of other users' likes and dislikes without our assistance during Phase 3. \added{Participants regarded the} like-dislike ratio as a significant influence on the recommendation algorithm. P5 stated that a small number of likes for a video with millions of views is a signal that \textit{``something is wrong''}. 

\subsubsection{The ``Others' Comments'' Influence}

Participants also suggested that the comments made by other users could signal to the \added{YouTube's recommendation system} that a specific video is relevant for the recommendations. P10, for example, suggests that the number of comments is a potential factor of influence. P12 considers that YouTube can use the comments done by other users \textit{``to measure the level of importance of a specific video''}. Likewise, P8 stated that commenting will increase a video's hypothetical rating.

\added{Here, again, participants did not elaborate on how the recommender system would analyze the comments technically. This lack of detail suggests that this influence is more related to social aspects, e.g., the number of comments that the algorithm considers to determine if a video is relevant enough to be recommended.}

\subsection{Recommender System Beliefs}

This group of beliefs relates to the inner workings of recommendation \added{systems} that do not \added{consider a} user's actions \added{on} the platform or the activity of other users \added{for the recommendations}. \added{The influences in this group are notable because they represent an intuitive grasp of basic concepts behind user-based, content-based, and context-aware recommendation systems. In this context, it is interesting to consider that these influences were not related to the interface, but suggested based on how users perceive and experience their recommendations.}

\subsubsection{The ``Who is Similar?'' Influence}

This factor relates to how the recommender system computes the similarity between users~(P12, P16, P10). P10, for example, \added{thought} that users with similar interests influence the recommendations. P16 explained that \textit{``there are, of course, other people watching similar content''} and \added{the algorithm is considering their tastes to provide similar recommendations to the participant}. Likewise, P12 believed that the recommendation process is finding people with similar tastes: 

\begin{quote}
They look for people that are similar to you, for the kind of taste that a person has [...] from there they also relate to other kinds of videos.
\end{quote}

\added{Notably, while the results indicate a certain level of awareness that the recommender algorithm considers the similarity between users, the interviewees were not able to explain how the recommendation system computes this similarity and what data it uses for this calculation. In addition to that, none of the participants elaborated on what makes a user similar to each other.}

\subsubsection{The ``What is Similar?'' Influence}

Participants also talked about \added{how the recommendation system analyzes the similarity} between content. In the context of music recommendations, P17 said: \textit{``I guess [it is] according to the music genre that I use for listening (...) they pick similar options to recommend [to] me, according to what I am searching for. They make a connection from it''}. P7 explained: \textit{``If I look at something in the sports section, I might have more suggestions about sports in general''}. P14 mentioned a similar effect in the opposite direction: \textit{``they have a database, and when they notice that you do not use those topics, then they stop suggesting them''}. 

\added{Others suggested that the recommendation system is based on how frequently a video is} viewed by users. P4 said: \textit{``The strongest factor, I would say, is quantity. If I always call up a topic now, then I'm sure that this will also be predominantly indicated to me''}. 

\added{Here, again, users showed a superficial level of awareness about how the recommender system considers content similarity, but they did not demonstrate an understanding of how the recommender algorithm analyzes the content of the videos. They also did not elaborate on how the topics in videos can be detected and how the comparison of different videos based on topics would work.}

\subsubsection{The ``When do I watch?'' Influence}

Participants expressed that time and date influence video recommendations~(P4, P10, P11) and that YouTube can use this information to make a distinction between home and work~(P4, P14). 

For instance, P4 described the \added{recommender system} as \textit{``thinking''}: \textit{``Attention, it is now 9 p.m., the boy is probably off work''}. P14 expressed how the recommendation system takes his context into account, \added{offering him music during working hours} and child\added{ren} videos at home to entertain his kid. Furthermore, P10 mentioned that the \added{recommendation system also} considers seasonal aspects such as recommending different music during Christmas time.

This belief is noteworthy in that it closely connects to research on context-aware recommendation systems. Academics have been researching such context-aware recommendation systems that take time, mood, or social context into account for some time~\cite{Adomavicius:2005:ICI:1055709.1055714, Gantner:2010:FMC:1869652.1869654, Liu:2010:ANM:1869652.1869653}, even though the systems have not gained widespread adoption yet.

\subsubsection{The ``Where do I watch?'' Influence}

Respondents mentioned the location of a user as another factor that influences the recommendation. According to the responses, location influences recommendations by providing content that is as close as possible to a user's current location and languages. P3 believes that: \textit{``Of course, they show you videos regarding your [own] language''}. P2 \added{said}: 

\begin{quote}
I'm quite sure that location will probably also be part of it [the recommendations]. It would be a little bit strange maybe, but strange in the sense that it can maybe recommend the closest nearby videos, I think.
\end{quote}

As a Costa Rican, P17 believed that the location \added{influences the recommendations because of the language used in her videos}, saying: \textit{``Sure. Because the majority of suggestions that I get are in Spanish, they almost do not show me anything in English''}. 

Interestingly, not all users agreed that a user's location influences the recommendations. P7 and P8 expressed similar opinions: ``No, I do not think that is so important for them. So, actually, it is not important at all.''

\subsection{Company Policy Beliefs}

\added{In addition to influences related to individual actions, social media dynamics, and the algorithms}, this group of beliefs references the influence of the specific decisions made by the organization that operates the \added{recommender system. Participants centered this group of beliefs on how individual choices made by YouTube as a company can influence their experience with the recommendations. This group highlights the responsibility and agency of the organization hosting the recommendation system.}

\subsubsection{The ``Paid Recommendation'' Influence}

A small group of participants expressed that some people could be paying for a position in the video recommendations~(P1, P8, and P10). P10, for instance, said: \textit{``I kind of think that you can buy places on YouTube''}. 

Interestingly, participants connected this belief to negative emotions. P8 expressed his disappointment: \textit{``I think it has to do with money and that is a pity''}. \added{Notably, this factor implies that users do not believe that YouTube and its recommendation system are neutral, and could be guided by commercial interest rather than offering neutral recommendations. Unlike other influence factors that are acting in the best interest of users by making recommendations more relevant, this belief implies that the recommender system is not genuinely calculating what the user prefers, but instead acting based on the agenda of the company that runs it.}

\subsubsection{The ``Third-Party Data-Sharing'' Influence}

Respondents also believed data-sharing practices between different companies, e.g., Google, Facebook, Instagram, and Twitter, influence the recommendations on YouTube~(P5, P13, and P11). P9\added{, for instance,} stated that the recommendations take information from other sources \added{into account}. 

\added{Several} participants commented on data-sharing \added{practices} between Alphabet Inc. and its subsidiaries, Google and YouTube. Participants described Google as \textit{``tracking everything''}~(P2) or as a \textit{``big data universe''}~(P9) that is collecting information to create YouTube recommendations. P9 believes that the system uses \textit{``every click, every search, every information''}. 

Surprisingly, respondents did not comment on the legality of such data-sharing practices. \added{The participants did not mention any existing data} protection laws, which are especially strong in the E.U. countries Belgium and Germany.

\subsubsection{The ``Psychological Experts'' Influence}

In the interviews, participants referred to the existence of psychological experts that work for YouTube and make decisions that influence the recommendations~(P7, P8). According to P7, YouTube is performing \textit{``psychological data mining''}, where \textit{``a whole team of psychologists''} tries to keep the users on the platform to make \textit{``as much profit as possible''}~(P7). This influence evoked strong negative emotions. P7 stated: 

\begin{quote}
It makes me feel sad about the world to know there is a whole team of psychologists [...] just [to] keep them [the users] and have as much profit as possible.
\end{quote}

\added{This factor represents a distrust in the video recommendations provided by YouTube, which participants regard as primarily aimed at generating profit for the company. Furthermore, this finding potentially connects to a lack of awareness and understanding of how such recommendation systems generate video recommendations.}

\section{Discussion}

User beliefs \added{about} video recommendation systems are co-produced through user interaction and the complex socio-technical system that generates the recommendations. For those who study this interaction and design these systems, it is essential to understand which aspects of this co-production are accessible to users, which are not, and how this dynamic promotes specific understandings of the system. 

With this paper, we investigated video consumer's beliefs about algorithmic recommendations on YouTube, the most widely used video recommendation system in the world at the time of the investigation. We zoomed in on the user beliefs of middle-aged video consumers with no technical or computer science background \added{and examined} a diverse participant pool \added{from} different countries. 

\added{Our investigation provides a variety of contributions. First, based on the analysis, we present a framework to distinguish the varied users' understandings of video recommendations based on the four main actors identified by video consumers. This framework brings design suggestions that could improve the experience with recommendations. The framework also includes a previously unexplored actor that affects recommendations: the organization that operates the system. Second, even though the consulted population did not grow up with social media or algorithmic systems, we report a high level of awareness of the recommendation system in the consulted population, in contrast to prior similar studies (e.g.,~\cite{Eslami2015,hamilton_path_2014}). However, video consumers' understanding is still very superficial. Third, we connect our results to similar studies, which suggest that our findings are generalizable without age concerns.}

\subsection{Four Actors That Influence User's Video Recommendations}

\begin{figure}%[h]
  \centering
  \includegraphics[width=\linewidth]{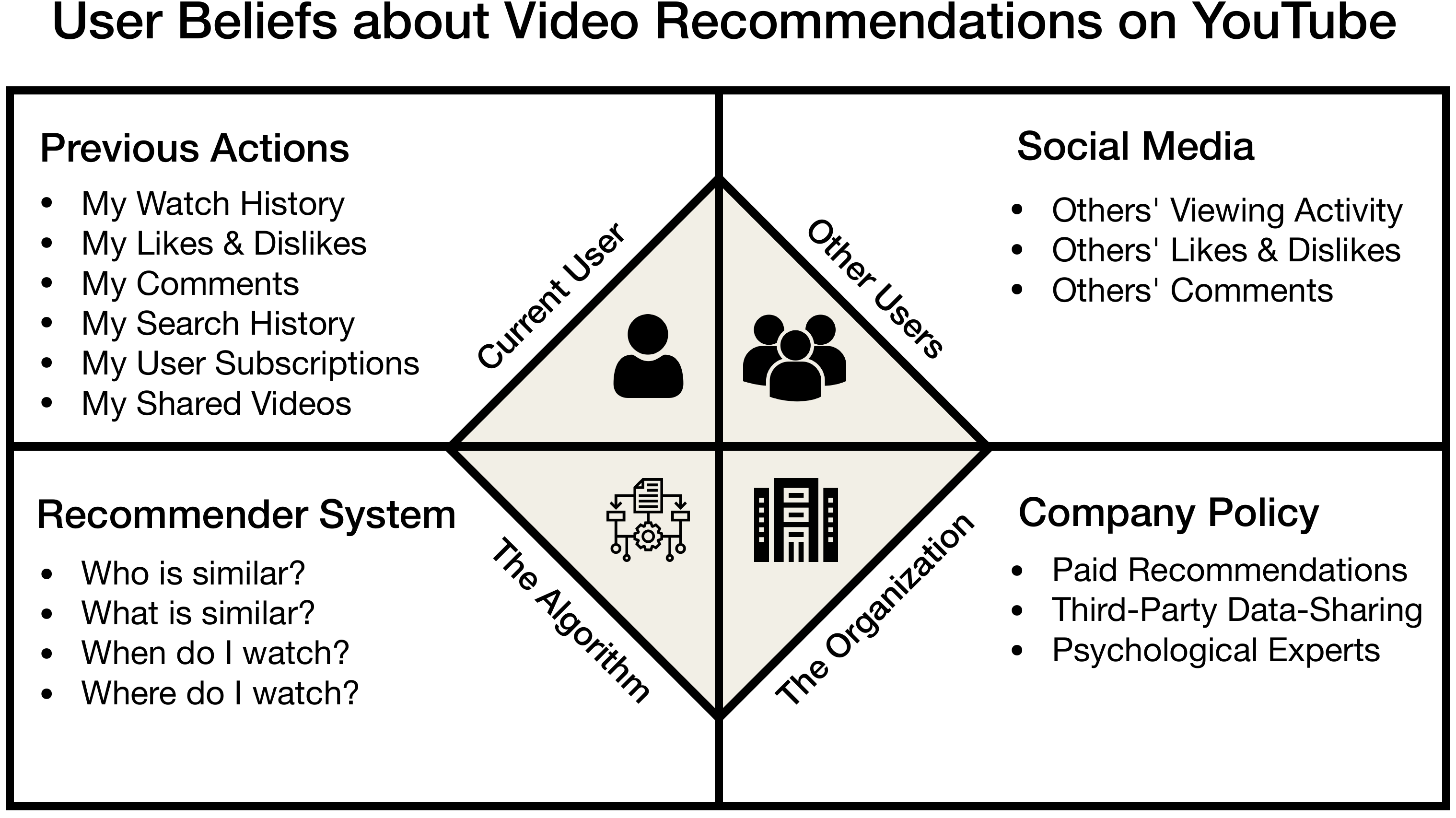}
  \caption{The Four Actors that Influence Video Recommendations and User Beliefs on YouTube.}~\label{fig:user_beliefs}
\end{figure}

Our thematic analysis identified four groups of user beliefs: Previous Actions, Social Media, Recommender System, and Company Policy. \added{These user beliefs can also be grouped based on the four main actors that influence} the recommendations: 1)~the current user, 2)~other users, 3)~the recommender system and 4)~YouTube, or the organization hosting the recommender system. Figure~\ref{fig:user_beliefs} \added{provides an overview of} the user beliefs and their related influence factors, grouped by these main actors.  

This distinction based on actors provides a better understanding of the \added{relationship} between user beliefs and the main actors that \added{users recognized as significant influences for their} recommendations. \added{It also} offers a framework to analyze \added{previously unexplored different user beliefs in recommendation systems. For instance, this} framework provides four main actors who could \added{serve as a departure point to inspire design suggestions to improve the experience of recommendation systems. Likewise, this framework enables} designers \added{and} scholars to envision \added{different} ways of shaping user beliefs about the system based on what is accessible and understandable for users. \added{This framework is based on empirical findings provided by middle-aged YouTube consumers with no strong background in technology, a previously unexplored population. Furthermore, the framework includes the organization that hosts the recommendation system as an actor that has a direct influence on the experience with the recommendations. New lenses based on these four actors' agency can provide design suggestions and serve as a starting point for future work to understand the socio-technical context of algorithmic recommendations.}

\added{For instance, a variety of the influence factors described in the analysis could be explained by a lack of technical understanding. Using the four categories recognized by the users in our framework, explanations could show whether the algorithm, a user's actions, other users' actions, or the organization that runs the platform is responsible for the presented recommendations. In the following}, we describe other possible \added{design opportunities based on} the four identified groups of beliefs and actors: 

\subsubsection{Previous Actions Belief, The Current User as an Actor}

This group of beliefs centers on the actions performed by the current user. These actions inform a user model, \added{which, in turn, is used to produce the video recommendations}. Examples for the actions we identified are liking, sharing, commenting, subscribing, or searching for a video. \added{Consequently,} designers should explore ways of visualizing the influence that past behavior or current actions of an individual user have on his or her video recommendations. 

It is important to note that all of the influence factors related to this belief are \added{mostly related to} interactive user interface elements. For designers of algorithmic systems, this implies that they have to pay special attention to how the interface presents these interactive elements because all of them \added{will} influence the user beliefs about the algorithmic system. For instance, even if the search bar does not have an explicit relation \added{to} the recommendations, \added{as our analysis showed,} users tend to relate their recommendation results \added{to} this interface element. 

Moreover, \added{design} efforts could encourage users to leverage these interactive elements to influence the recommender system intentionally. For instance, designers could add control features to allow users to review and correct how their watching history influences their personalized recommendation\added{s}. \added{This design suggestion is supported by previous research that found that explanations are a useful tool to give users a sense of control over the results produced by a recommendation system~\cite{Rader2018}.}

\subsubsection{Social Media Belief, Other Users as Actors} 

This group of beliefs references the actions of other users, i.e., all the other people who are also using \added{the recommendation system}. \added{This activity is} commonly measured via statistics \added{such as} views, likes, dislikes, and comments \added{to identify popular or peer-related content for the recommendations}.

In this context, the actions of other users and their influence \added{on} the recommendations could be made more transparent. One suggestion for designers is to visualize how and to what degree actions of other YouTube users influence the video recommendations. Designers could also add user interface elements that allow the user to control how much other users' actions influence their recommendations. \added{As shown in our results, participants believe they regularly receive unwanted video recommendations because of the popularity of specific content with other users. Therefore, enabling users to control this influence could be useful to achieve a better experience with recommendations. Design suggestions and recommendations from previous studies on} social media platforms could be adapted to video recommender systems to improve their algorithmic experience~\cite{Alvarado2018}.

\subsubsection{Recommender System Belief, The Algorithm as the Actor}

This belief relates to the influence of the recommender systems \added{and its technical implementation}. Here, the algorithm is \added{an} actor recognized by \added{our} participants. 

\added{This view of} the algorithm as an actor that influences the recommendations connects to related work on algorithmic personas~\cite{Wu:2019:AGD:3371885.3359321} \added{and algorithms as social-vigilant entities~\cite{doi:10.1177/2053951720923377}}. \added{Interestingly, this connection implies that both YouTube producers, passive video consumers, and users of other recommendations systems recognize the direct influence that the algorithm as a technical actor has on their experience with video recommendations, although with specific differences.}

According to our participants, the algorithm influences the video recommendations \added{in} two dimensions: similarity and context. \added{On the one hand,} users referenced the similarity between users and the similarity between videos. These influence factors directly connect to the technical inner workings of recommender systems, which are commonly item-based or user-based collaborative filtering~\cite{Jannach:2016:RSB:3013530.2891406, Jugovac:2017:IRR:3143523.3001837}. \added{On the other hand, }the context influence described the time and place where the user watches a video, informing a user model that relates to the context in which the algorithm creates the recommendation. In contrast to other influence factors, context can be ephemeral, i.e., it can change from minute to minute. Incorporating context into recommendation is\added{, therefore,} a challenging task.

\added{A possible} design recommendation informed by this belief is to provide explanations of how the recommendation system calculates the similarities between users or videos. \added{As expressed by our participants, the interface could tell whether the system base a recommendation on the similarity between previously watched videos, or similarities with other users.} It should also be transparent how date, time, context, and location are taken into account when making a recommendation. \added{As expressed by our participants, the platform could make it transparent whether the recommender system is offering videos dedicated to work or leisure time.}

\subsubsection{Company Policy Belief, The Organization as the Actor}

\added{Our participants believed that the organization that provides the recommendation system directly impacts their experiences with video recommendations.} Participants reflected on how YouTube as an organization \added{could} influence their recommendations \added{based on corporate decisions. As described by our participants,} two influences in this \added{area} (Paid Recommendations and Psychological Experts) \added{were} associated with strong negative emotions and experiences. 

Therefore, \added{it seems practitioners should} also consider how decisions made in an organization influence \added{user beliefs}. Users recognize these influence factors, build beliefs towards them, and will configure their use of a system based on the role the organization plays in making recommendations. \added{Additionally}, algorithmic system designers and organizations as a whole should \added{consider this aspect and} further \added{investigate} what is leading to those negative emotions \added{, how users form these beliefs, how designers can control them, and what is needed to promote a better experience with recommender systems in this area}. 

\added{Based on the results in our analysis,} this group of user beliefs should motivate organizations to \added{be} transparent \added{with decisions such as paid} positions in the recommendations, if the organization shares data with third parties, or whether the organization uses psychological experts or experiments to increase engagement with the platform. These actions could be studied to determine whether they increase trust, acceptability, and improve the experience with video recommendations.

\added{General suggestions for this area include a closer, more productive, and more transparent communication regarding such decisions. Finally, it is worth mentioning to promote the implementation of principles derived from laws like the EU General Data Protection Rules, which could improve users' experience within this area. In contrast to similar studies that did not explicitly point out the organization as an actor, these results open a novel space to consider specific design recommendations in the context of algorithmic recommendations.}

\subsection{Algorithmic Awareness of the Recommendation System on YouTube}

For the participants in our investigation, \added{we found that a superficial form of} algorithmic awareness of YouTube's recommendation system was comparatively high. More than 89\% of participants (16 out of 18) were aware \added{of the algorithmic recommendations}. The two participants who did not know about the recommendations recognized them after they were pointed out by the researchers. This reality means that the vast majority of this sample, with no technical or computer science background, was aware of YouTube's recommendations. We also found that the majority of participants~(72\%) actively use YouTube's recommendations. 
\added{These results contrast with prior} work that investigated algorithmic awareness~\cite{hamilton_path_2014, Eslami2015}. Earlier studies on Facebook \added{showed that users might not be aware of recommendation systems.} 

\added{However, participants' understanding of the inner workings of the algorithm remains limited and vague. Participants were not able to understand or explain how the system works in detail and refer to the influence factors in general terms, thinking that something like sharing a video ``has consequences''~(P4) or that the viewing activity of others ``creates a tendency''~(P6). While participants are, to some extent, able to articulate beliefs about ``what'' is influencing their recommendations, but they are not able to explain ``how'' it influences their recommendations. This finding suggests that their understanding of the recommendations is superficial and limited.}

\added{This superficial level of understanding of the recommendations could be explained by recent media reports that} have increased the level of awareness about algorithmic systems \added{without increasing understanding}. Another possible reason could be the prominent and familiar role that the recommendations play in the user interface of YouTube and the fact that the system clearly labels the recommendations. \added{This does not keep users from developing in(accurate) beliefs, as we will discuss in the following section.}

\added{Our users recognized a diverse range of influence factors that explain how they think the recommendation system on YouTube works. This large number of possibilities expressed by our participants indicates that there is no explicit agreement regarding how middle-aged video consumers without a background in technology think such systems works. This finding suggests that, while a superficial awareness of a system is high, the understanding of how it works is comparatively low.}

\added{This could be connected to a} lack of transparency of such recommendation system\added{s}. In the context of Facebook's News Feed, Rader and Gray described a potential feedback loop in which \added{1)~user beliefs about a platform can influence users' behavior, which 2)~potentially affect the input to an algorithm, 3)~which, in turn, could influence the output of the algorithm, which 4)~again can affect user beliefs}~\cite{Rader2015}. This reality directly connects to newspaper articles that increasingly argue that YouTube's recommendation algorithm acts as a catalyst for filter bubbles and online ``radicalization''~\cite{roose_YouTubes_2019, Lewis2018,nytimes_YouTube_radicalizer_2018,buzzfeed_las_vegas_2017,nytimes_chemnitz_2018}.

Recognizing the different user beliefs around recommendation systems is an essential first step towards addressing these issues. Moreover, understanding how users reason about complex algorithmic curation systems can motivate further research to make influence factors more visible to users. 

\subsection{(In)Accuracy of (Un)Intuitive User Beliefs}
% what is publicly known about YT, comparison

The following section serves two purposes. On the one hand, we explore the (in)accuracy of the user beliefs. On the other hand, we explore which of these user beliefs are intuitive, i.e., which did come up without further assistance and \added{which} were only agreed to by the participants. 

Since we have no affiliation with YouTube, we cannot conclusively assess how accurate our users' beliefs are. We situate our findings by relating them to what YouTube has made publicly known about the system. Technical papers published by YouTube and public information available to software developers offer insights into the recommendation system\added{'s inner workings}. This knowledge allowed us to compare the beliefs expressed by our participants to official statements by YouTube. 

Influences such as ``My Watch History'', ``Who is similar?'' and ``What is similar?'', and ``My Search History'' were all mentioned by our participants without assistance, which means that they come ``naturally'' to users. ``My Comments'' and ``Other's Comments'' as well as ``My Likes \& Dislikes'' and ``Others' Likes \& Dislikes'', were not brought up by the participants. Respondents did, however, believe that they influenced recommendations when we mentioned these influence factors.

As personal explanations, all beliefs and influence factors are valid and have merit for the individual and for researchers that want to understand users' perspectives of complex recommendation algorithms. Accordingly, the goal is not to check which beliefs are true or false but to use these user beliefs as a lens to understand what aspects of user recommendations are understandable and accessible to middle-aged \added{video consumers} without high ICT literacy. \added{Here, it is worth noting that recognition may be easier than recall, following Nielsen's general principles for interaction design~\cite{nielsen199510}. However, this difference could indicate that some influence factors are more ``natural'' to users than others. Meanwhile, the difference between recognition and recall effects needs further research.}

We found that participants did not mention a variety of aspects that are known to influence the video recommendations. Examples of unmentioned influences include demographics, video titles, video descriptions, thumbnails, co-watching, or time spent watching a particular video. Video thumbnails were also never mentioned as a factor of influence on the presentation of the recommended videos. \added{Participants} did not refer to \added{the influence of the playlists} and whether \added{the system should include} some videos in the recommendation\added{s} because they are newsworthy, i.e., \added{apart from the personalized recommendations}. \added{One} explanation for not mentioning these aspects could be a lack of general knowledge about machine learning \added{techniques} or natural language processing.

\subsection{Comparing Influences to Previous Studies}

We further situate the results of this study by comparing them to previous similar efforts on user beliefs and folk theories. \added{It is worth noticing the similarities among our results and previous work considering our focus on middle-age users.}

Influences such as ``My Watch History'', ``Others' Viewing Activity'', ``My Comments'', ``Others' Comments'', ``My User Subscriptions'', and ``My Shared Videos'' are similar to the Global Popularity Theory found by Eslami et al.~\cite{Eslami2016}. The Global Popularity Theory represents the belief that the number of likes and comments primarily measures the likelihood of content appearing in Facebook's News Feed. Eslami et al. also formulated the Narcissus theory, a theory related to the ``What is similar?'' influence in our study, which states that the similarity to a friend is a strong influence factor. Likewise, the Eye of Providence theory, which describes Facebook as having a God-like all-seeing eye watching over users, can be related to the ``When do I watch'', ```Where do I watch'' and the ``Psychological Experts'' influences~\cite{Eslami2016}. All of these influences are also similar to the operational theories found by DeVito et al.~\cite{DeVito2017}. 

Influence factors such as ``My Watch History'', ``My Search History'', ``My Comments'', ``Others' Comments'', ``My User Subscriptions'', ``My Likes \& Dislikes'', and ``Others' Likes \& Dislikes'', are specific to YouTube as a platform. Even if previous studies do not consider them explicitly, other investigations seem to reflect similar findings, e.g., Alvarado et al.'s study of the algorithmic experience of movie recommendations~\cite{Alvarado2019}, in which users ask for a better understanding and more control over the influence of all of these and similar interaction opportunities with the video recommender algorithms. Likewise, explorations of algorithmic experience in movie recommendations~\cite{Alvarado2019} and social media~\cite{Alvarado2018} portray similar results related to the ``Third-Party Data-Sharing'' influence.

\subsection{Current Research On User Beliefs about Algorithmic Recommendations}

It is noteworthy that three of the four \added{groups of} beliefs we identified \added{can be related to} academic conferences \added{that frequently address topics related to the different user beliefs. These conferences include} the ACM Conference on User Modeling, Adaptation and Personalization (Previous Actions), the ACM Conference on Hypertext and Social Media (Social Media), \added{and} the ACM Recommender Systems Conference (Recommender System). While it is reassuring that academic conferences address these user beliefs, \added{the expertise is scattered across different communities}. We argue that a more holistic approach that considers all four groups of user beliefs \added{would be valuable} to understand the socio-technical context of video recommendations. 

\added{Meanwhile, there is no dedicated conference that explores how the organization influences the experience with the recommender system (Company Policy)}. \added{Possibly, this is an area of research that the CSCW community could focus on to achieve a holistic understanding of the complex socio-technical nature of algorithmic recommendations.}

\section{Limitations and Future Work}

The main goal of this study was to understand user beliefs about YouTube's recommendation algorithm held by middle-aged video consumers without high ICT literacy. Two-thirds of the participants were between 37 and 43 years old, limiting our \added{findings' generalizability} for other age groups. While we made sure that we recruited from countries with high levels of YouTube usage, the generalizability of our findings beyond these countries is hypothetical. The video consumers in our sample were highly educated, which could limit the generalizability of our findings. The vast majority of participants were aware of the recommendations before the investigation, which means that we do not know how well users with less algorithmic awareness can reason about YouTube's algorithm. 

Our results show that the use of specific features of the platform affected the \added{participants'} different influences. For instance, a difference exists between users who comment regularly and those who do not comment at all. Unfortunately, the study design did not allow us to examine such differences systematically. Future work could examine the implications that use practices have on user beliefs and how different the user beliefs of other user groups are. Another practice was that many participants stated that they rarely or never use the \textit{``Like''} buttons. \added{Also}, at least five users reported that they do consume YouTube a lot without having a personal account. It would be valuable to study how this practice influences user beliefs, algorithmic awareness, and the experience with the recommendations \added{in general}.

A significant fraction of the participants in our investigation also mentioned another practice: they share their accounts with other people, particularly younger children. During the interviews, participants continuously expressed how these practices influence their understanding of the algorithm. This finding echoes similar studies about Netflix~\cite{Alvarado2019}, in which the algorithmic experience of the movie recommendation system is affected by \textit{algorithmic social practices}.

Finally, this study focused on YouTube, which could have produced results that are unique to this platform. Further research is needed to discover whether the results can be generalized to other video recommendation services, streaming services, or movie recommendations.

We invite more researchers to examine user beliefs of algorithmic video recommendation quantitatively. Future work with a non-purposive, representative sampling could investigate how frequent\added{ly} each of the different user beliefs \added{occurs} and which of the different user beliefs co-occur with each other.

\section{Conclusion}

This paper presents the first overview of middle-aged video consumers' beliefs about YouTube's algorithmic video recommendations. The analysis identified four groups of user beliefs that describe how users understand video recommendations on YouTube: Previous Actions, Social Media, Recommender System, and Company Policy. For each user belief, we identified several influence factors \added{recognized by users which} could help designers improve the experience of video recommendations. 

\added{To enable solutions to this problem, we systematically analyzed the different influence factors and identified four dominant user beliefs that we relate to different actors. Users recognize these influences without having a background in technology. We invite further studies on the complex socio-technical context of recommender systems using this framework, including how the organization that operates the recommendation system influences the user experience with the recommendations.}

%%
%% The acknowledgments section is defined using the "acks" environment
%% (and NOT an unnumbered section). This ensures the proper
%% identification of the section in the article metadata, and the
%% consistent spelling of the heading.
\begin{acks}
Part of this research has been supported by the Deutsche Forschungsgemeinschaft (DFG, German Research Foundation) -- Projektnummer 374666841 -- SFB 1342, the KU Leuven Research Council (grant agreement C24/16/017), and the University of Costa Rica (UCR).  We thank all participants for their time and their insights. We also thank the reviewers for their helpful and constructive feedback. The icons are by Pixel perfect from www.flaticon.com.
\end{acks}

%%
%% The next two lines define the bibliography style to be used, and
%% the bibliography file.
\bibliographystyle{ACM-Reference-Format}
\bibliography{references,oscarrefs}

%%
%% If your work has an appendix, this is the place to put it.
%\appendix

\end{document}